\documentclass[preprint,12pt]{elsarticle}

\usepackage{amssymb}

\newcounter{bla}

\journal{Journal of Computational Science}

\begin{document}

\begin{frontmatter}



\title{OpenSBLI: A framework for the automated derivation and parallel execution of finite difference solvers on a range of computer architectures}


\author[a]{Christian T. Jacobs\corref{cor1}}
\author[a]{Satya P. Jammy}
\author[a]{Neil D. Sandham}

\cortext[cor1] {Corresponding author.\\\textit{E-mail address:} C.T.Jacobs@soton.ac.uk}
\address[a]{Aerodynamics and Flight Mechanics Group, Faculty of Engineering and the Environment, University of Southampton, University Road, Southampton, SO17 1BJ, United Kingdom}

\begin{abstract}
Exascale computing will feature novel and potentially disruptive hardware architectures. Exploiting these to their full potential is non-trivial. Numerical modelling frameworks involving finite difference methods are currently limited by the `static' nature of the hand-coded discretisation schemes and repeatedly may have to be re-written to run efficiently on new hardware. In contrast, OpenSBLI uses code generation to derive the model's code from a high-level specification. Users focus on the equations to solve, whilst not concerning themselves with the detailed implementation. Source-to-source translation is used to tailor the code and enable its execution on a variety of hardware.
\end{abstract}


\end{frontmatter}


\section{Introduction}\label{sect:introduction}
High Performance Computing (HPC) systems and architectures are evolving rapidly. Traditional single processor-based CPU clusters are moving towards multi-core/multi-threaded CPUs. At the same time new architectures based on many-core processors such as graphics processing units (GPUs) and Intel's Xeon Phi are emerging as important systems and further developments are expected with energy-efficient designs from ARM and IBM. According to the IT industry, such advances are expected to deliver compute hardware capable of exascale-performance (i.e. $10^{18}$ floating-point operations per second) by 2018 \citep{Thibodeau_2009}. Yet many frameworks aimed at computational/numerical modelling are currently not ready to exploit such new and potentially disruptive technologies.

Traditional approaches to numerical model development involve the production of static, hand-written code to perform the numerical discretisation and solution of the governing equations. Normally this is written in a language such as C or Fortran that is considerably less abstract when compared to a near-mathematical domain specific language. Explicitly inserting the necessary calls to MPI or OpenMP libraries enables the execution of the code on multi-core or multi-thread hardware. However, should a user wish to run the code on alternative platforms such as GPUs, they would likely need to re-write large sections of the code, including calls to new libraries such as CUDA or OpenCL, and optimise it for that particular hardware backend \citep{Rathgeber_etal_2012}. As HPC hardware evolves, an increasing burdon faced by computational scientists becomes apparent; in order to keep up with trends in HPC, not only must a model developer be a domain specialist in their area of study, but also an expert in numerical algorithms, software engineering, and parallel computing paradigms \citep{JacobsPiggott_2015, Rathgeber_etal_Submitted}.

One way to address this issue is to introduce a separation of concerns using high level abstractions, such as domain specific languages (DSLs) and active libraries \citep{Rathgeber_etal_Submitted, LoggWells_2010, Logg_etal_2012, Alnaes_etal_2014, Luporini_etal_2015}. This paradigm shift allows a domain specialist to describe their problem as a high-level, near-mathematical specification. The task of taking this specification and transforming it into executable computer code can then be handled in the subsequent abstraction layer; unlike the traditional approach of hand-writing the C/Fortran code that discretises the governing equations, this layer generates the code automatically from the problem specification. Finally, the generated code can be readily targetted towards a specific hardware platform through source-to-source translation. Hence, domain specialists focus on the equations they wish to solve and the setup of their problem, whilst the parallel computing experts can introduce support for new backends as they become available. At no point does the code have to undergo a fundamental re-write if the desired backend changes. Use of such strategies can have significant benefits for the productivity of both the user and developer, by removing the need to spend time re-writing code and/or the problem specification \citep{LoggWells_2010}.

Given the motivation for the use of automated solution techniques, in this paper we present a new framework, OpenSBLI, for the automated derivation and parallel execution of finite difference-based models. This is an open-source release of the recent developments in the SBLI codebase developed at the University of Southampton, involving the replacement of SBLI's Fortran-based core with flexible Python-based code generation capabilities, and the coupling of SBLI to the OPS active library \citep{Giles_etal_2015, Reguly_etal_2014, Mudalige_etal_2014, Jammy_etal_2015} which targets the generated code towards a particular backend using source-to-source translation. Currently, OpenSBLI can generate OPS-compliant C code to discretise and solve the governing equations, using arbitrary-order central finite difference schemes and a choice of either the forward Euler scheme or a third-order Runge-Kutta time-stepping scheme. OpenSBLI then uses OPS to produce code targetted towards different backends. It is worth noting that backend APIs such as OpenMP (version 4.0 and above) are also capable of running on CPU, GPU and Intel Xeon Phi architectures, for example. However, currently OPS has no support for OpenMP version 4.0 and above. Moreover, codes that are written by hand in OpenMP would still potentially need to be re-written if different algorithms or equations were to be considered. Thus, the benefits of code generation still play a crucial role here, regardless of which backend is chosen.

The application of SBLI has so-far concentrated on problems in aeronautics and aeroacoustics, in particular looking at shock-boundary layer interactions (see e.g. \cite{TouberSandham_2009, DeTullioSandham_2010, Redford_etal_2012, Wang_etal_2015} and the references therein for more details). While such applications entail solving the 3D compressible Navier-Stokes equations, in principle other equations expressible in Einstein notation and solved using finite differences are also supported by the new code generation functionality, highlighting another advantage of such a flexible approach to numerical model development. Note also that while OpenSBLI does not yet feature shock-capturing schemes and Large Eddy Simulation models (unlike the legacy SBLI code), these will be implemented in the future as part of the project's roadmap. The main purposes of this initial release is the algorithmic changes to legacy SBLI's core.

Details the abstraction and design principles employed by OpenSBLI are given in Section \ref{sect:design}. Section \ref{sect:verification_validation} details three verification and validation test cases that were used to check the correctness of the implementation. The paper finishes with some concluding remarks in Section \ref{sect:conclusion}.

\section{Design}\label{sect:design}
Legacy versions of SBLI comprise static hand-written Fortran code, parallelised with MPI, that implements a fourth-order central differencing scheme and a low-storage, third or fourth-order Runge-Kutta timestepping routine. It is capable of solving the compressible Navier-Stokes equations coupled with various turbulence parameterisations (e.g. Large Eddy Simulation models) and diagnostic routines. In contrast, OpenSBLI is written in Python, and by replacing the legacy core with modern code generation techniques, the existing functionality of SBLI is enriched with new flexibility; the compressible Navier-Stokes equations can still be solved in OpenSBLI for the sake of continuity, but the set of equations that can be readily solved essentially becomes a superset of that of the legacy code. Furthermore, the use of the OPS library allows the generated code to easily be targetted towards sequential, MPI, or an MPI+OpenMP hybrid backend (for CPU parallel execution), CUDA and OpenCL (for GPU parallel exection), and OpenACC (for parallel execution on accelerators), without the need to re-write the model code. OPS is readily extensible in terms of new backends, making the code generation technique an attractive way of future-proofing the codebase and preparing the framework for exascale-capable hardware when it arrives. The main achievement of OpenSBLI is the ability to express model equations at a high-level with the help of the SymPy library \citep{SymPy_2016}, expanding the equations based on the index notation, and coupling this functionality with the generation of OPSC-based model code and also with the OPS library which performs code targetting. OpenSBLI's focus on the generation of computational kernels essentially forms a bridge between the high-level equations and the computational parallel loops (`parloops') that iterate over the grid points to solve the governing equations.

\begin{figure}[!ht]
   \begin{center}
      \includegraphics[width=0.9\columnwidth]{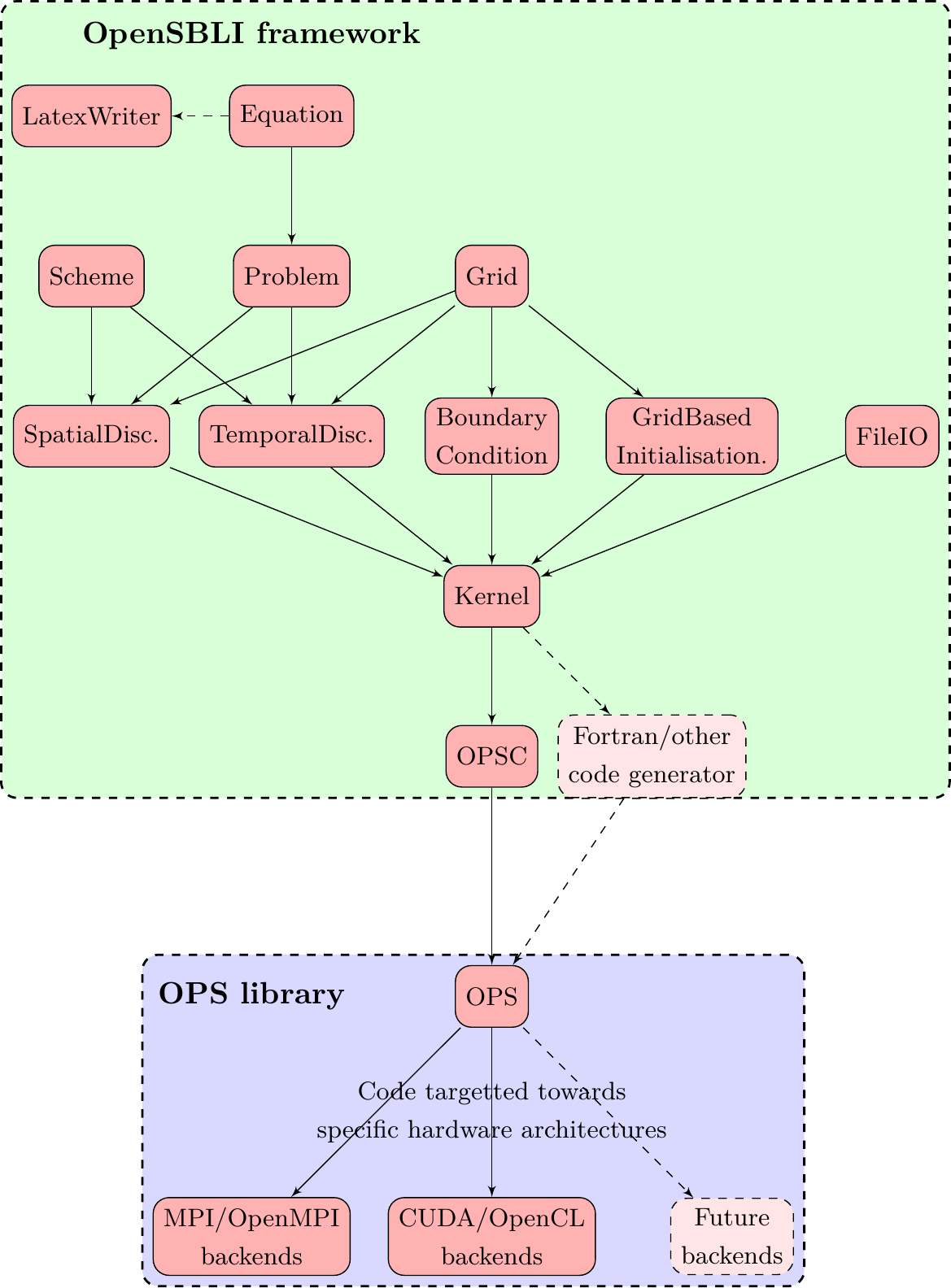}
      \caption{The overall design of the OpenSBLI framework with respect to the core classes. The code targetting happens within the OPS library. The CPU backends include MPI, OpenMP, hybrid MPI+OpenMP, as well as a sequential version of the code. The GPU backends include CUDA and OpenCL, which can also be combined with MPI to run the code on multiple GPUs in parallel. The only accelerator backend available is OpenACC.}
      \label{fig:design}
   \end{center}
\end{figure}

For any given simulation that is to be performed with OpenSBLI, the problem (comprising the equations to be solved, the grid to solve them on, their associated bondary and initial conditions, etc) must be defined in a setup file, which is nothing but a Python file which instantiates the various relevant components of the OpenSBLI framework. All components follow the principle of object-oriented design, and each class is explained in detail throughout the subsections that follow. An overview of the class relationships is also provided in Figure \ref{fig:design}.

\subsection{Equation specification}
In a similar fashion to other problem solving environments such as OpenFOAM \citep{OpenFOAM_2014}, Firedrake \citep{Rathgeber_etal_Submitted}, FEniCS \citep{LoggWells_2010, Logg_etal_2012}, OPESCI-FD \citep{Sun_2016}, Devito \citep{Kukreja_etal_2016, Lange_etal_2016}, deal.II \citep{Bangerth_etal_2007} and FreeFEM++ \citep{Hecht_2012}, OpenSBLI comprises a high-level interface for specifying the differential equations that are to be solved. These equations (and any accompanying formulas for temperature-dependent viscosity, for example) can be expressed in Einstein notation, also known as index notation. The adoption of such an abstraction is advantageous since it removes the need for the user to expand the equations by hand which can be an error-prone task. Furthermore, much like the Devito domain specific language (DSL) \citep{Kukreja_etal_2016, Lange_etal_2016} for finite difference stencil compilation, OpenSBLI makes use of the SymPy symbolic algebra library that supplies the basic components required for the modelling functionality that has been implemented in the present work. This functionality includes the automatic expansion of indices based on their contraction structure, such that repeated indices are expanded into a sum about that index, and the implementation of various types of differential operator.

\subsubsection{Expressing}
Consider the conservation of mass equation

\begin{equation}\label{eq:mass_pre_expansion}
  \frac{\partial \rho}{\partial t} + \frac{\partial}{\partial x_j}\left[\rho u_j\right] = 0,
\end{equation}
where $u_j$ is the $j$-th component of the velocity vector $\mathbf{u}$, $\rho$ is the density field, and $x_j$ is the coordinate field in the $j$-th dimension. In an OpenSBLI problem setup file, the user would specify this as a string, giving the left-hand side and right-hand side of the equation in the following format:

\texttt{mass = "Eq(Der(rho, t), -Conservative(rho*u\_j, x\_j))"}

The functions \texttt{Der} and \texttt{Conservative} here are OpenSBLI-specific derivative operators, each defined in their own class derived from SymPy's \texttt{Function} class. Other high-level interfaces such as OpenFOAM offer similar differential operators such as \texttt{div} and \texttt{grad}, for example \citep{OpenFOAM_2014}. General derivatives are represented using the \texttt{Der} operator, whereas the \texttt{Conservative} operator ensures that the derivative will not be expanded using the product rule. A skew-symmetric form of the derivative is also available using the \texttt{Skew} function, discussed later in Section \ref{sect:taylor_green_vortex}. All of these are essentially `handler'/placeholder objects that OpenSBLI uses for spatial/temporal discretisation after parsing and expanding the equations about the Einstein indices. Special functions such as the Kronecker delta function and the Levi-Civita symbol are also available, derived from SymPy's \texttt{LeviCivita} and \texttt{KroneckerDelta} classes in order to handle Einstein expansion; these too are expanded later by OpenSBLI.

\subsubsection{Parsing}
Once all of the governing equations have been expressed by the user in string format, they are collected together in OpenSBLI's \texttt{Problem} class (see \ref{sect:setup}). This class also accepts \textit{substitutions}, \textit{formulas}, and \textit{constants}. For long equations, such optional \textit{substitutions} (such as the definition of the stress tensor) can be written as a separate string (in the same way as the governing equations) to allow better equation readability, and then automatically substituted into the equations (such as the conservation of momentum and energy equations) at expansion-time instead of performing such error-prone manipulations by hand. The constitutive equations which define a relationship between the prognostic and non-prognostic variables are given as \textit{formulas}, for example temperature-dependent viscosity relations, and an equation of state for pressure. The \textit{constants} are the spatially and temporally independent variables which are represented as strings. Upon instantiation of the \texttt{Problem} class, the process is invoked to transform the equations into their final expanded form.

For each equation in string form, a new OpenSBLI \texttt{Equation} object is created. During its initialisation, SymPy's \texttt{parse\_expr} function converts the equation string into a SymPy \texttt{Eq} data type. Any of the OpenSBLI derivative operators such as \texttt{Der} and \texttt{Conservative} (currently in string format) are replaced by actual instances of the \texttt{Der} and \texttt{Conservative} classes. Similarly, any substitutions given in the \texttt{Problem} are parsed and substituted directly into the expression using SymPy's \texttt{xreplace} function. All other terms in the parsed expression are represented by OpenSBLI's \texttt{EinsteinTerm} class, derived from SymPy's \texttt{Symbol} class, which contains its own methods and attributes for determining/expanding Einstein indices. For example, the class's initialisation method \texttt{\_\_init\_\_} splits up the term \texttt{u\_j} where there are underscore markers, and stores the Einstein index \texttt{j} in a list as a SymPy \texttt{Idx} object. The \texttt{get\_expanded} method later replaces the alphabetical Einstein indices with actual numerical indices, replacing \texttt{\_j} with \texttt{0} and \texttt{1}, in the 2D case. Finally, any constants in the \texttt{Problem} object are also represented as an \texttt{EinsteinTerm} object, but are flagged as constant terms in OpenSBLI, so that they are not spatially or temporally-dependent. The coordinate vector components $x_j$ (and the time term $t$) are a special case of an \texttt{EinsteinTerm}; these are marked with a \texttt{is\_coordinate} flag so that, during the expansion phase, the \texttt{EinsteinTerm}s are made dependent on the coordinate field (and time, if appropriate) to ensure that differentiation is performed correctly.

\subsubsection{Expanding}
After the parsing and substitution stage, the equations are expanded about repeated indices. Note that this process is performed by OpenSBLI, although various SymPy classes underpin the functionality. Following the example, (\ref{eq:mass_pre_expansion}) would be expanded as

\begin{equation}\label{eq:mass_post_expansion}
  \frac{\partial{\rho}}{\partial{t}} + \frac{\partial}{\partial x_0}\left[\rho u_0\right] + \frac{\partial}{\partial x_1}\left[\rho u_1\right]= 0.
\end{equation}

OpenSBLI loops over each \texttt{EinsteinTerm} stored in the parsed \texttt{Equation} object, and maps it to a SymPy \texttt{Indexed} object. For example, the term \texttt{u\_k} would first be mapped to \texttt{u[k]}. The index \texttt{k} in the term is then expanded over 0, $\ldots$, $d-1$ (where $d$ is the dimension of the problem) by replacing it with each integer dimension, yielding a SymPy \texttt{MutableDenseNDimArray} array of size $d$ (for a vector function, or $d \times d$ for a tensor of rank 2) of expanded variables which is stored as a class attribute. For example, expanding the vector \texttt{u[k]} yields the expansion array \texttt{[u0, u1]} in 2D. Upon expansion, the terms are also made spatially-dependent (i.e. indexed by $x_0$, $x_1$, $x_2$ coordinates, depending on the dimension) and, if applicable, temporally-dependent (i.e. indexed also by $t$). The only exceptions to this are constants such as the Reynolds number $\mathrm{Re}$. The expansion array from the previous example then becomes \texttt{[u0[x0, x1, t], u1[x0, x1, t]]} (and \texttt{[x0, x1]} for the constant coordinate field).

Each equation is expanded by locating any repeated indices and then summing over them as appropriate. For example, after mapping each \texttt{EinsteinTerm} (e.g. \texttt{u\_k}) to an \texttt{Indexed} object (e.g. \texttt{u[k]}), the mass equation is represented internally as

\texttt{Eq(Der(rho, t), -Conservative(rho*u[k], x[k]))}

Since the index \texttt{k} is repeated, the expansion arrays are used to expand this expression to

\texttt{Eq(Der(rho[x0, x1, t], t), -Conservative(rho[x0, x1, t]*u0[x0, x1, t], x0[x0, x1, t]) - Conservative(rho[x0, x1, t]*u1[x0, x1, t]), x1[x0, x1, t]))}

Finally, the \texttt{Der} and \texttt{Conservative} functions are applied, with the expression becoming

\texttt{Eq(Derivative(rho[x0, x1, t], t), -Derivative(rho[x0, x1, t]*u0[x0, x1, t], x0) - Derivative(rho[x0, x1, t]*u1[x0, x1, t], x1))}

\noindent which is equivalent to (\ref{eq:mass_post_expansion}). Similar expansion can also be applied for any other equations involving e.g. diagnostic fields. Note how the calls to \texttt{Der} and \texttt{Conservative} have been replaced by calls to SymPy's \texttt{Derivative} class (which in turn uses SymPy's \texttt{diff} function); while it is SymPy that handles the differentiation, it is OpenSBLI that handles the exact formulation of the derivative (i.e. OpenSBLI has ensured that the derivative has not been expanded using the product rule here).

Any nested derivatives are also handled here. It is not currently possible to specify, for example, \texttt{diff(diff(u\_j, x\_i), x\_j}) using SymPy's \texttt{diff} function directly because the fact that \texttt{u\_j} is dependent on \texttt{x\_i} and \texttt{x\_j} is not taken into account. In contrast, the use of \texttt{Der} and \texttt{EinsteinTerm}s like \texttt{u\_j} in OpenSBLI allows the derivative to be computed correctly since the terms are made dependent through the use of \texttt{Indexed} objects as previously described. OpenSBLI users must instead use the \texttt{Der} function \texttt{Der(Der(u\_j, x\_i), x\_j)}. For each nested derivative (or nested function in general), the inner function is evaluated first along with all other non-nested functions. Only then is the outer function applied.

For the purposes of debugging, OpenSBLI includes a \texttt{LatexWriter} class that takes the expanded equations as input and writes them out in LaTeX format so developers can more easily spot errors, for example where indices have been expanded incorrectly.

\subsection{Grid}
\begin{figure}[!ht]
   \begin{center}
      \includegraphics[width=0.48\columnwidth]{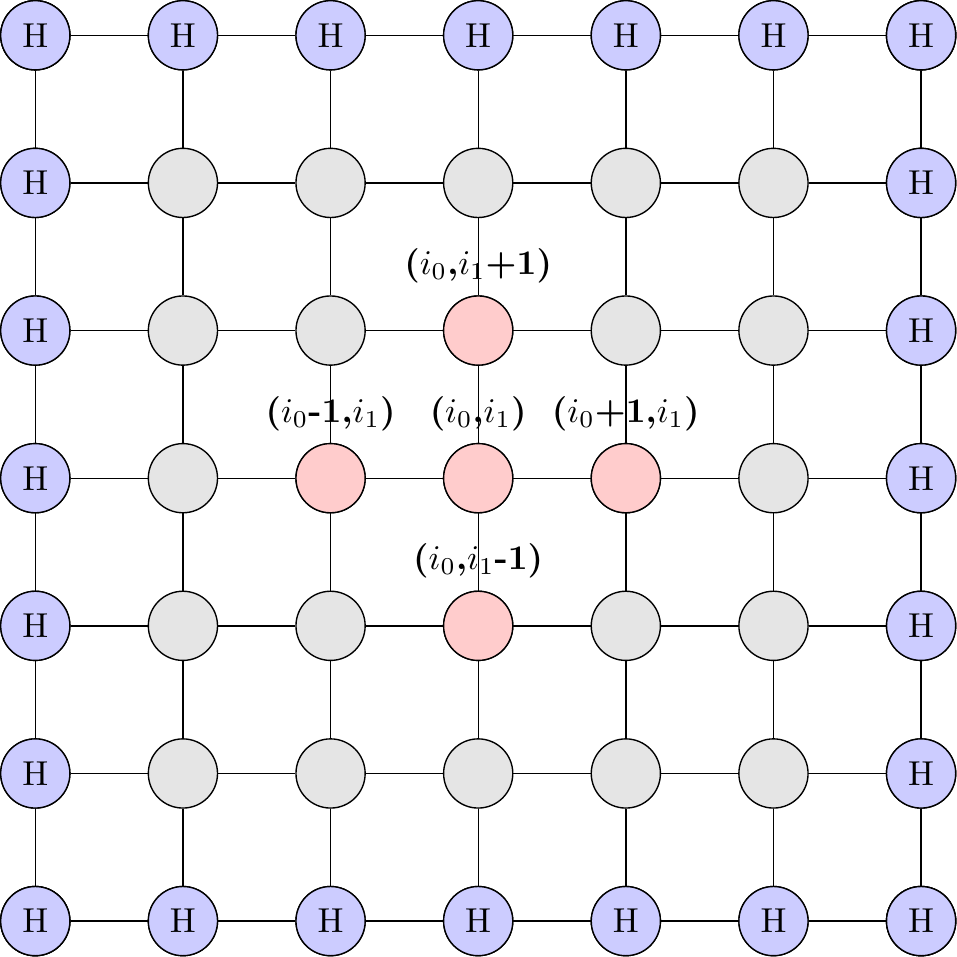}
      \caption{The regular grid of solution points upon which the governing equations are solved. The grid point indices in the $x$ and $y$ directions are denoted $i_0$ and $i_1$, respectively. The halo points that surround the outer-most points of the domain are labelled `H'. A computational stencil used for second-order central differencing is highlighted red in the center, with the relative grid coordinates of each point.}
      \label{fig:grid}
   \end{center}
\end{figure}

The governing equations are discretised on a regular grid of solution points that span the domain of interest; an example is provided in Figure \ref{fig:grid}. All grid-related functionality is handled by the \texttt{Grid} class, which must be instantiated by the user in the problem setup file. The dimensionality of the problem $d$, the number of points in each dimension, and the grid spacing must all be supplied. A problem of dimension $d$ would generate a grid of $N_{x_0} \times \ldots \times N_{x_{d-1}}$ solution points in total, where $N_{x_i}$ represents the user-defined number of grid points in direction $x_i$.

For the sake of looping over each solution point and computing the necessary derivatives via the finite difference method, each (non-constant) term is processed further by OpenSBLI; the index of each spatial coordinate (e.g. \texttt{x0}) is mapped onto an index over the grid points in that spatial direction (e.g. \texttt{i0}) which will iterate from 0 to $N_{x_i}-1$ (for a given direction $x_i$) when the computational kernel is eventually generated.

In addition to the solution points within the physical domain, a set of halo points (or `ghost' points), which border the outer-most grid points, are also created automatically depending on the boundary conditions and the spatial order of accuracy. These halo points are necessary to ensure that the derivatives near the boundary can be computed with the same stencil as the `inner' points. The exact number of halo points required therefore depends on the number of stencil points; for example, in Figure \ref{fig:grid} the stencil for a second-order central difference (using 3 points in each direction) would require one halo point at each end of the domain. The values that these halo points hold depend on the type of boundary condition applied, and this is discussed in more detail in Section \ref{sect:bcs}.

Every field/term in the governing equations that is represented by the grid indices holds a so-called `work array' which essentially contains the field's numerical value at each of the grid points, including the halos. The implementation of initial and boundary conditions is done by accessing and modifying this work array, as will be described in Sections \ref{sect:ics} and \ref{sect:bcs}.

\subsection{Computational kernels}\label{sect:kernels}
The \texttt{Kernel} class defines a sequence of computational steps that should be performed to solve the governing equations. For instance, one kernel may be created to compute the spatial derivative of a field, while another kernel handles the initialisation of the field values based on a given initial condition, and another handles the enforcement of boundary conditions that involve computations. During the instantiation of a kernel, the relevant variables and fields are classified as inputs, outputs and input/outputs (i.e. both an input and an output), and the kernel's range of evaluation (i.e. the range of grid indices over which the kernel is applied). This helps to minimise data transfer, since only those variables/fields required to perform the computation are passed to the generated kernel code.

\subsection{Discretisation schemes}
Once a grid is created, the equations are discretised upon that grid. For spatial discretisation purposes OpenSBLI offers a central differencing scheme for first and second-order derivatives; all the stencil coefficients are computed using SymPy, which allows stencils of an arbitrary order of accuracy to be created. For temporal discretisation purposes, OpenSBLI features the (first-order) forward Euler scheme as well as the same low-storage, third-order Runge-Kutta timestepping scheme \citep{CarpenterKennedy_1994} present in the legacy SBLI code.

To use a particular scheme, one should instantiate a discretisation scheme derived from the generic base class called \texttt{Scheme}, which essentially stores the finite difference stencil coefficients or the weights used in a particular time-stepping scheme. Spatial and temporal schemes should be instantiated separately.

For the purpose of spatial discretisation, handled by the OpenSBLI \texttt{SpatialDiscretisation} class, an \texttt{Evaluations} object is created for each of the formulas, and the derivatives in the equations. Each \texttt{Evaluations} object automatically finds and stores the dependencies of a given term (e.g. $\partial (A + B)/\partial x_0$ requires the dependencies $A$ and $B$). Once all the \texttt{Evaluations} have been created, they are sorted with respect to their dependencies being evaluated (e.g. if $B$ depends on $A$, then $A$ should be evaluated first). The next step involves defining the range of grid point indices over which each evaluation should be performed, and also assigning a temporary work array for each evaluation. All of the evaluations are then described by a \texttt{Kernel} object (see Section \ref{sect:kernels}). It is here, while creating the kernels, that the (continuous) spatial derivatives are automatically replaced by their discrete counterparts. It should be noted that, for the evaluation of formulas, these kernels are fused together if they have no inter-dependencies to avoid race conditions when running on threaded architectures. Finally, to evaluate the residual for the purposes of temporal discretisation, the derivatives in the expanded equations (represented by an \texttt{Evaluations} object) are substituted by their temporary work arrays, and a \texttt{Kernel} is created for evaluating the residual of each equation.

The temporal discretisation, handled by the \texttt{TemporalDiscretisation} class, involves applying the various stages of the time-stepping scheme supplied using the residuals computed by the spatial discretisation process. Similarly, a \texttt{Kernel} object is created for the evaluations in the time-stepping scheme.

\subsection{Initial conditions}\label{sect:ics}
In order for the prognostic fields to be advanced forward in time, initial conditions can be applied using the \texttt{GridBasedInitialisation} class. This is accomplished in much the same way as specifying equations, but involves assignment of grid variables and work arrays of grid point values. For example, in the simulation setup file the $x_0$ coordinate can be defined using the grid point index and $\Delta x_0$:

\texttt{x0 = "Eq(grid.grid\_variable(x0), grid.Idx[0]*grid.deltas[0])"},

\noindent which in turn defines the initial value for each prognostic variable, by assigning this to the array of values at each grid point (also known as the variable's work array), e.g.:

\texttt{rho = "Eq(grid.work\_array(rho), 2.0*sin(x0))"}.

\subsection{Boundary conditions}\label{sect:bcs}
OpenSBLI currently comprises two types of boundary condition, implemented in the classes \texttt{PeriodicBoundaryCondition} and \texttt{SymmetryBoundaryCondition}. Users may apply different boundary conditions in different directions if they so wish. Periodic boundaries are defined such that, for each prognostic field $\phi$, $\phi(x_0) = \phi(x_N)$ where $N$ is the number of points in the domain. This condition is achieved via the exchange of halo point data at each end of the domain. Symmetry boundary conditions enforce the condition that $\phi(x_N) = \phi(x_{N-1})$ for scalar fields and $\phi_i(x_N) = -\phi_i(x_{N-1})$ for vector fields (in the direction $i$), which is achieved using a computational kernel.

\subsection{Input and output}
The state of the prognostic fields can be written to disk every $n$ iterations as defined by the user, or only at the end of the simulation. This functionality is handled by the \texttt{FileIO} class. OpenSBLI adopts the HDF5 format \citep{FolkPourmal_2010, Collette_2013} as it features parallel read/write capabilities and therefore has the potential to overcome the serial input/output bottleneck currently plaguing many large-scale parallel applications \citep{Padua_2011, Fu_etal_2011}. Future releases of OpenSBLI will come with the ability to read in mesh files and the state fields from an HDF5 file, enabling the restarting of simulations from `checkpoints' as well as the assignment of initial conditions that cannot be simply defined by a formula.

\subsection{Code generation}
OpenSBLI currently generates code in the OPSC language which performs the simulation; this is essentially standard C++ code that includes calls to the OPS library. Such functionality is accomplished using the OpenSBLI \texttt{OPSCCodePrinter} class (derived from SymPy's \texttt{CCodePrinter} class, used to perform the generation of OPSC code statements) and the \texttt{OPSC} class (which agglomerates the literal strings of OPSC statements and kernel functions and writes them to file). The generated code's structure follows a generic template that maps out the order in which the simulation steps/computations are to be called. The template is represented as a multi-line Python string template, with each line containing a place-holder for the code that performs a particular step. Examples include \texttt{\$header} which is replaced by any generic boilerplate header code (e.g. \texttt{\#include <stdlib.h>} and kernel function prototypes), \texttt{\$initialisation} which is replaced by the grid and field setup (e.g. by declaring an OPS block using the \texttt{ops\_decl\_block} function), and \texttt{\$bc\_calls} which is replaced by calls to the boundary condition kernel(s). This template can be readily changed to incorporate additional functionality, such as the inclusion of turbulence models. Once all component place-holders have been replaced by OPSC code, the code is written out to disk. For the case of the OPSC language, two files are written; one is a C++ header file containing the computational kernels, and the other is the C++ source file containing various constant definitions (e.g. the timestep size \texttt{delta\_t}, and the constants of the Butcher tableau for the time-stepping scheme), OPS data structures, and calls to the kernels specified in the header file.

\begin{figure}[!ht]
   \begin{center}
      \includegraphics[width=0.95\columnwidth]{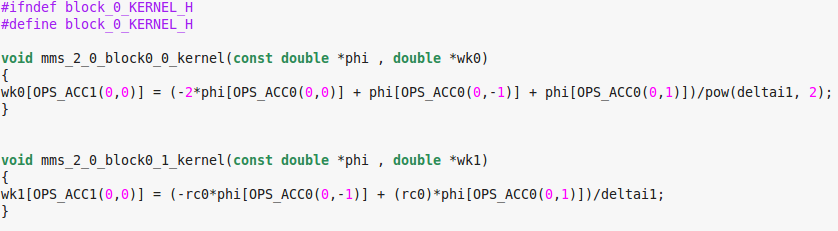}
      \caption{Code snippit showing two kernels from a 2D `method of manufactured solutions' (MMS) simulation (see Section \ref{sect:mms}) using second-order central differences. The first kernel computes $\frac{\partial^2 \phi}{\partial x_1^2}$ and stores it in a new work array called \texttt{wk0}. Similarly, the second kernel computes the first derivative $\frac{\partial \phi}{\partial x_1}$. The constant \texttt{deltai1} represents the grid spacing in the $x_1$ direction, and \texttt{rc0} holds the constant value of 0.5. Calls to \texttt{OPS\_ACC} are used to access the finite difference stencil structure.}
      \label{fig:kernels}
   \end{center}
\end{figure}

\begin{figure}[!ht]
   \begin{center}
      \includegraphics[width=0.95\columnwidth]{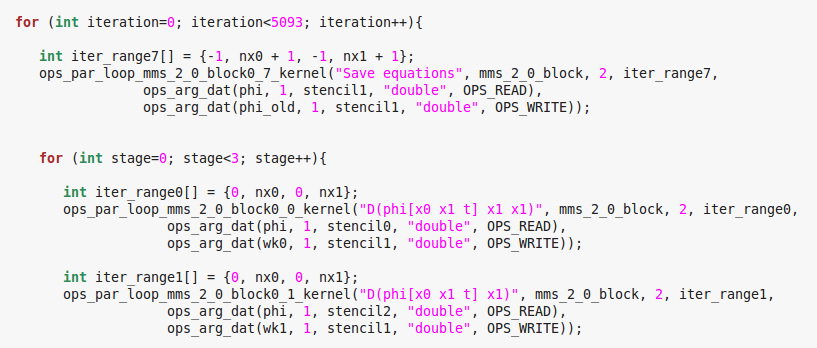}
      \caption{Code snippit showing the calls to the two kernels in Figure \ref{fig:kernels} inside the inner for-loop. The outer loop is the time-stepping loop which iterates for a user-defined number of iterations. The inner loop iterates over the 3-stages of the low-storage, third-order Runge-Kutta scheme. Note that each kernel is actually executed as an \texttt{ops\_par\_loop} over all the grid points.}
      \label{fig:kernel_calls}
   \end{center}
\end{figure}

OpenSBLI's local Python objects (most pertinently, the kernel objects that describe the computations to be performed on the grid) are essentially translated to OPSC data structures and function calls during the preparation of the code. For instance, when declaring computational stencils that define a particular central differencing scheme, the local grid indices stored in the \texttt{Central} scheme object are used to write out an \texttt{ops\_stencil} definition during code generation. Similarly, \texttt{ops\_halo} structures and calls to \texttt{ops\_halo\_transfer} are produced to facilitate the implementation of the periodic boundary conditions. All fields are declared as \texttt{ops\_dat} datasets; for an example of where these are used, see the function \texttt{ops\_argument\_call} in the file \texttt{opsc.py} which generates/accumulates calls to the OPS function \texttt{ops\_arg\_dat} through the use of `printf'-style string formatting, filling in the `placeholder' arguments (e.g. \texttt{\%s} in Python) with values from the local OpenSBLI objects. Finally, calls to OpenSBLI \texttt{Kernel} objects are represented in OPSC as regular C++ functions (see Figure \ref{fig:kernels}) which are passed to the \texttt{ops\_par\_loop} function (see Figure \ref{fig:kernel_calls}), which executes the function efficiently over the range of grid points within the desired block; OpenSBLI is currently a single-block code so only one block, containing all the grid points, is used. Further details on the OPS data structures and functionality can be found in the work by \cite{Reguly_etal_2014}.

Some optimisations are performed during the code generation stage by OpenSBLI to avoid unnecessary and expensive division operations in the kernels; rational numbers (e.g. finite difference stencil weights that are rational) and constant \texttt{EinsteinTerm}s raised to negative powers (e.g. $\mathrm{Re}^{-1}$) are evaluated and stored (e.g. by over-riding the \texttt{\_print\_Rational} method in the \texttt{OPSCCodePrinter} class).

Once the code generation process is complete, the OPS library is called to target the code towards various backends. These include the sequential code, MPI and hybrid MPI+OpenMP parallellised versions of the code for CPUs, CUDA and OpenCL versions of the code for GPUs, and an OpenACC version for accelerators. The test cases presented in this paper (see Section \ref{sect:verification_validation}) consider the sequential, MPI, and CUDA backends. Targetting `hand-written'/manually-generated model code towards a particular architecutre is something that is well-known as a time-consuming, error-prone and often unsustainable activity; often numerical models have to be completely re-written, involving many if-else statements and \texttt{\#ifdef}-style pragmas to ensure that the correct branch of the code is followed for a given backend. As the number of backends grows, the code becomes unsustainable. In contrast, with the abstraction introduced here through code generation, support for a new backend only needs to be added to the OPS library; the top-level, abstract definition of the equations and their implementation need not be modified due to the separation of concerns, thereby highlighting one of the key advantages of automated model development.

When comparing the number of lines and the complexity of the code that gets generated by OpenSBLI, another advantage of automated model development becomes clear; in the case of the 3D Taylor-Green vortex test case, the problem specification file containing $\sim$100 lines generates OPSC code that is approximately 1,500 lines long (excluding blank lines and comments). As more parameterisations (e.g. Large Eddy Simulation turbulence models) and diagnostic field computations are added, it is expected that this number would grow even further relative to the number of lines required in the setup file.

\section{Verification and Validation}\label{sect:verification_validation}
In order to verify the correctness of OpenSBLI and be confident in the ability of the solution algorithms to accurately represent the underlying physics, three representative test cases covering 1, 2 and 3 dimensions were created and are presented here.

\subsection{Propagation of a wave}
This 1D test case considers the first-order wave equation, given by

\begin{equation}\label{eq:wave}
\frac{\partial{\phi}}{\partial{t}} + c \frac{\partial \phi}{\partial x} = 0,
\end{equation}
where $\phi$ is the quantity that is transported at constant speed $c$. The expected behaviour is that an arbitrary initial profile at time $t=0$ is displaced by a distance $d_t = ct$, such that $\phi(x,t=0) = \phi(x=d_T,t=T)$ for some finish time $T$. The constant $c$ was set to 0.5 ms$^{-1}$ in this case, and the equation was solved on the line 0 $\leq x \leq$ 1 m. Eighth-order central differencing was used to discretise the domain in space in conjunction with a third-order Runge-Kutta scheme for temporal discretisation. The grid spacing $\Delta x$ was set to 0.001 m, and the timestep size $\Delta t$ was set to 4 $\times$ 10$^{-4}$ s, yielding a Courant number of 0.2. A smooth, periodic initial condition $\phi(x, t=0) = \sin(2\pi x)$ was used, and periodic boundary conditions were enforced at both ends of the domain.

\begin{figure}[!ht]
   \begin{center}
      \includegraphics[width=0.48\columnwidth]{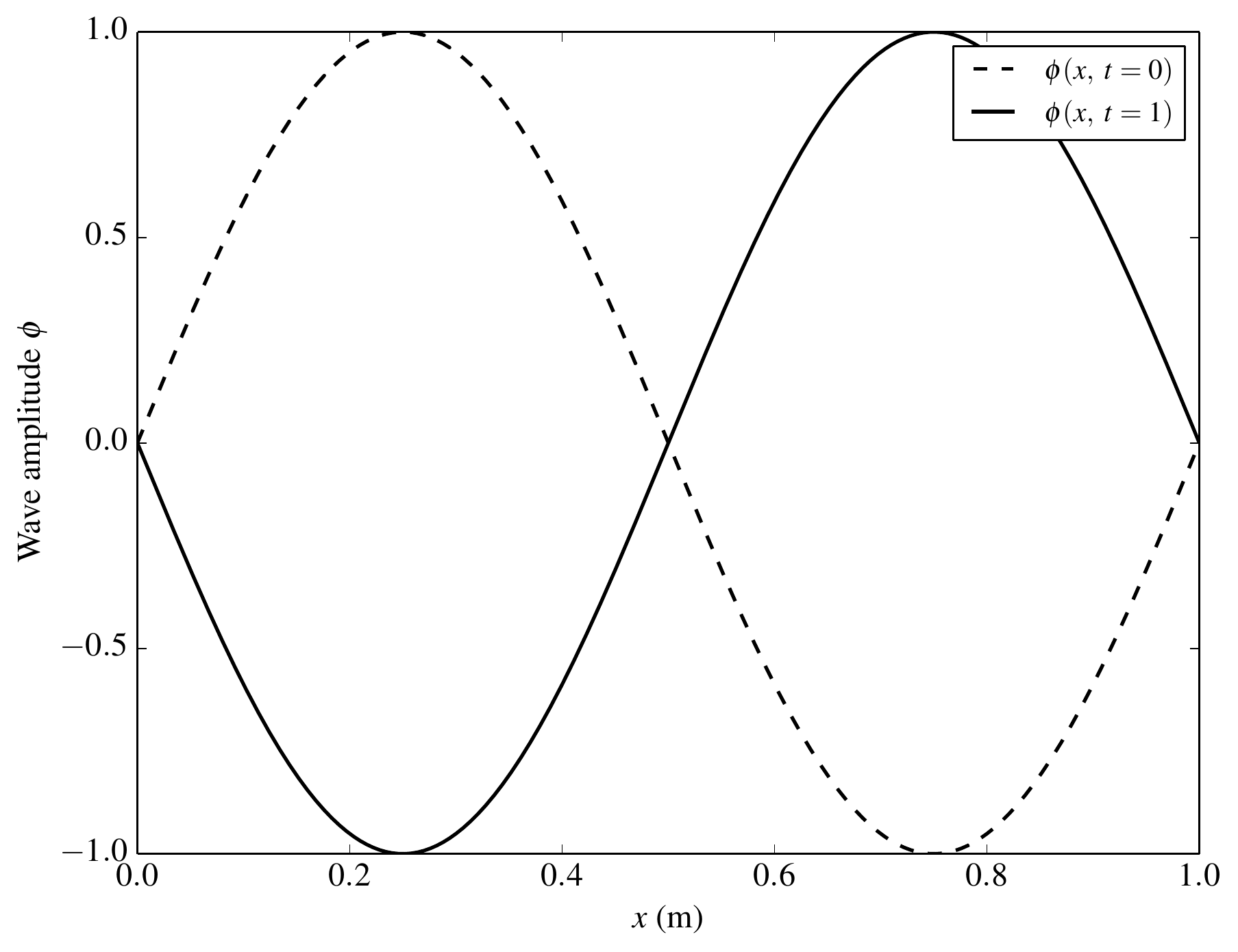}
      \includegraphics[width=0.48\columnwidth]{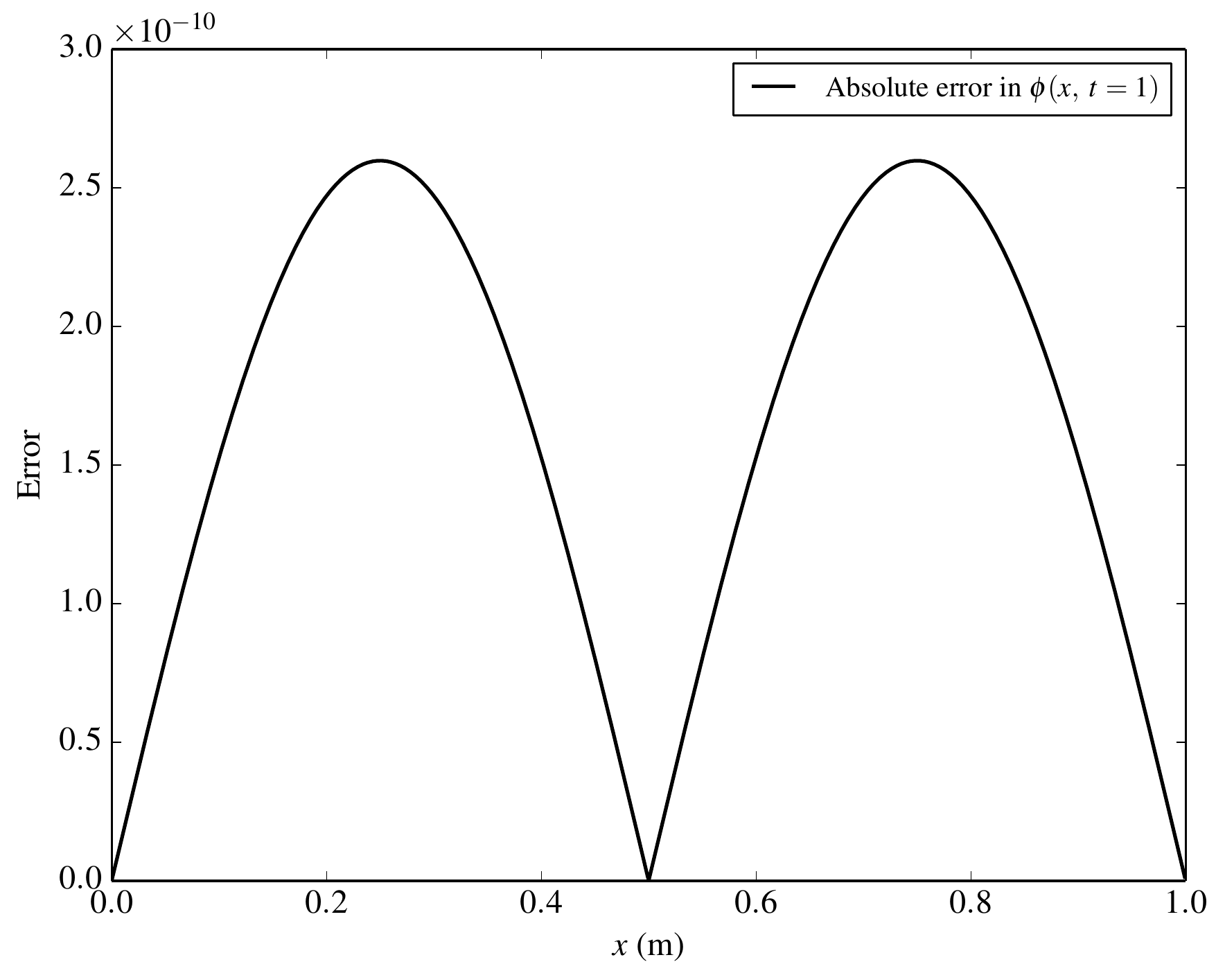}
      \caption{Results from the 1D wave propagation simulation. Left: The solution field $\phi$ at time $t$ = 0 s and $t$ = 1 s. Right: The error between the analytical solution and the numerical solution at time $t$ = 1 s.}
      \label{fig:wave_phi}
   \end{center}
\end{figure}

The simulation was run in serial (on an Intel\textregistered\  Core\texttrademark\  i7-4790 CPU) until a finish time of $t$ = 1 s. The initial and final states of the solution field $\phi$ are shown in Figure \ref{fig:wave_phi}. As desired, the error in the solution is very small at $O(10^{-10})$, and provides some confidence in the implementation of the solution method and the periodic boundary conditions.

\subsection{Method of manufactured solutions}\label{sect:mms}
The method of manufactured solutions (MMS) is a rigorous way to check the correctness of a numerical method's implementation \citep{SalariKnupp_2000, Roache_2002, Roy_2005}. The overall algorithm involves constructing a manufactured solution $\phi_m$ for the prognostic variable(s) $\phi$ and substituting this into the governing equation. Since the manufactured solution will not, in general, be the exact solution to the equation, a non-zero residual term will be present. This residual term is then subtracted from the RHS such that the manufactured solution essentially becomes the exact/analytical solution of the modified equation (i.e.~the one with the source term). A suite of simulations can then be performed using increasingly fine grids to check that the numerical solution converges to the manufactured solution at the expected rate determined by the discretisation scheme.

For this test, the 2D advection-diffusion equation (with a source term $S$) given by

\begin{equation}
   \frac{\partial \phi}{\partial t} + \frac{\partial}{\partial x_j}\left[\phi u_j - k\frac{\partial \phi}{\partial x_j}\right] + S = 0,
\end{equation}
is considered.

The constant $k$ is the diffusivity coefficient which is set to 0.75 m$^2$s$^{-1}$ here. The prescribed field $u_i$ is the $i$-th velocity component, with $u_0$ = 1.0 ms$^{-1}$ and $u_1$ = -0.5 ms$^{-1}$. The prognostic field $\phi$ is to be determined and has an initial condition of $\phi(x, t=0) = 0$. In a similar fashion to the works of \cite{SalariKnupp_2000, Roache_2002, Roy_2005}, the manufactured/`analytical' solution $\phi_m = \sin(x_0)\cos(x_1)$ employs a mixture of sine and cosine functions since these are continuous and infinitely differentiable. The SAGE framework \citep{SteinJoyner_2005} was used to symbolically determine the residual/source term $S$.

The domain is a 2D square with dimensions $0 \leq x_0 \leq 2\pi$ m and $0 \leq x_1 \leq 2\pi$ m such that the manufactured solution is periodic. Furthermore, periodic boundary conditions are applied on all sides of the domain. Six central differencing schemes of order 2, 4, 6, 8, 10 and 12 are considered for the spatial discretisation, and a third-order Runge-Kutta scheme is used throughout to advance the equation in time. To perform the convergence analysis, the grid spacing was halved for each successive case such that $\Delta x$ = $\Delta y$ = $\frac{\pi}{2}$, $\frac{\pi}{4}$, $\frac{\pi}{8}$, $\frac{\pi}{16}$ and $\frac{\pi}{32}$. The timestep size $\Delta t$ was also halved for each case to maintain a maximum bound of 0.025 on the Courant number; this was purposefully kept small and near-constant to minimise the influence of temporal discretisation error \citep{Vedovoto_etal_2011}. All simulations were run in serial (on an Intel\textregistered\  Core\texttrademark\  i7-4790 CPU) until a finish time of $T$ = 100 s to ensure that a steady-state solution was attained.

\begin{figure}[!ht]
   \begin{center}
      \includegraphics[width=\columnwidth]{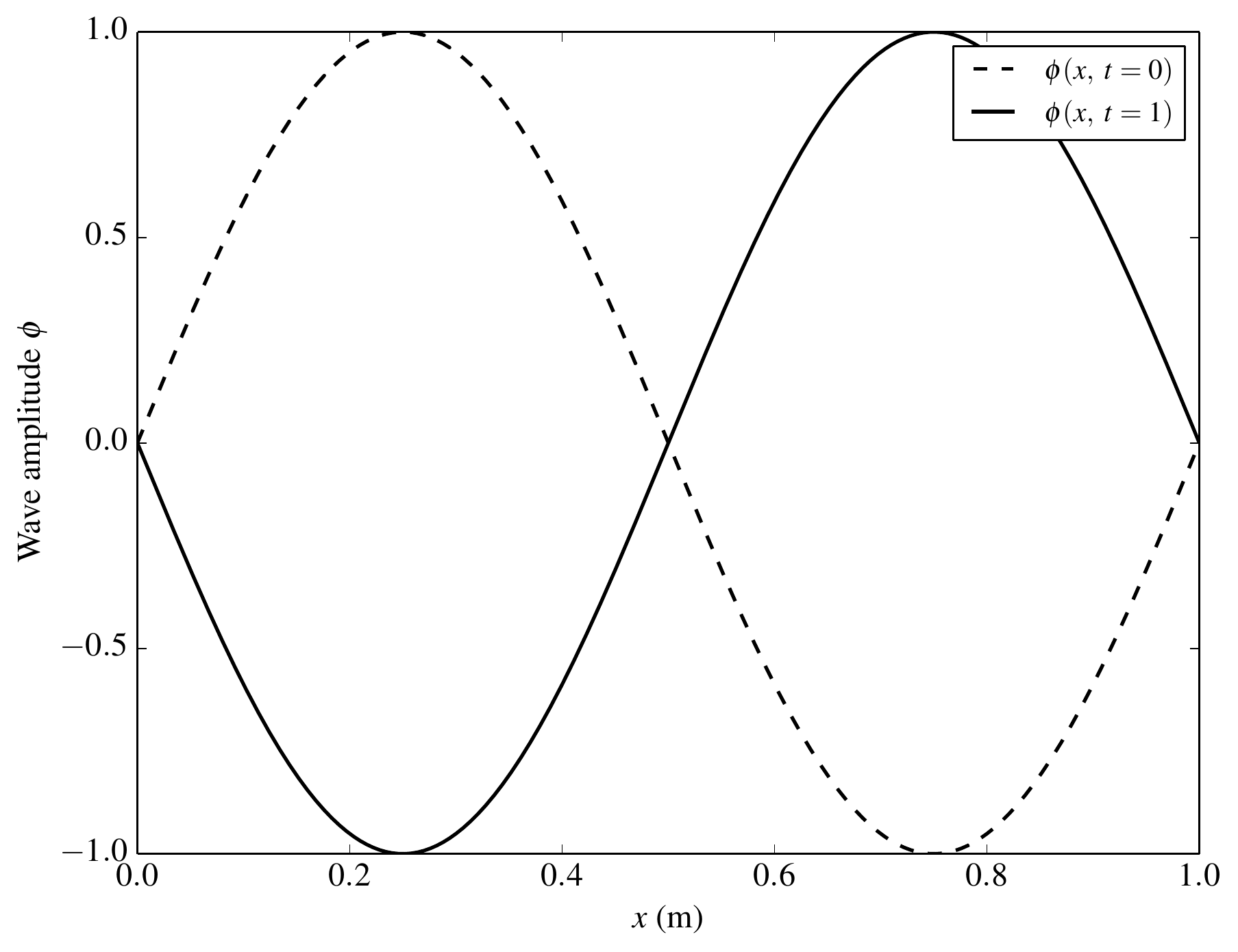}
      \caption{(a-e): The numerical solution field $\phi$ at the finish time $t = T$. (f): The manufactured solution $\phi_m$. All results are from the twelfth-order MMS simulation set.}
      \label{fig:mms_phi}
   \end{center}
\end{figure}

\begin{figure}[!ht]
   \begin{center}
      \includegraphics[width=\columnwidth]{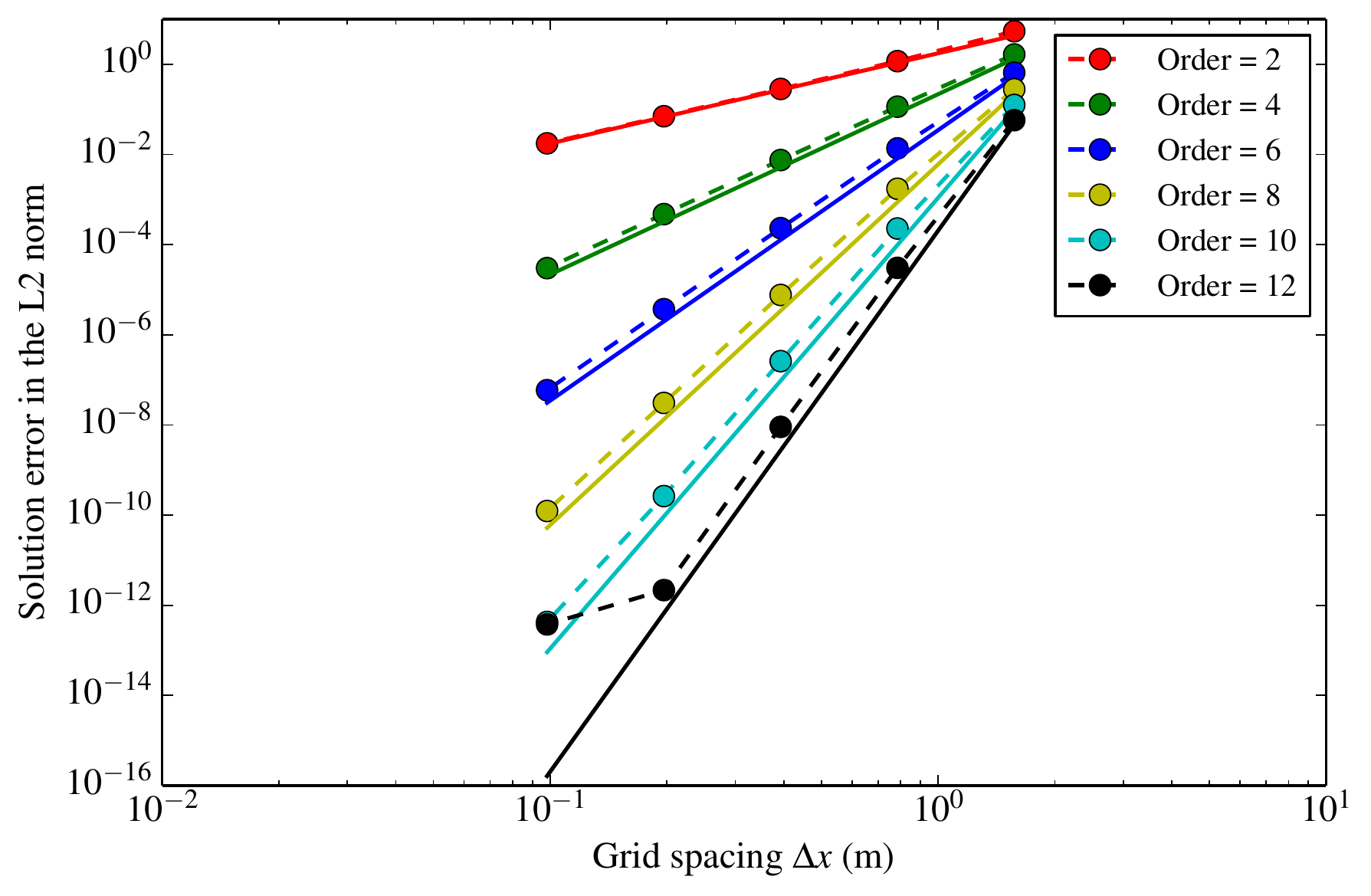}
      \caption{The absolute error (in the L2 norm) between the numerical solution $\phi$ and the exact/manufactured solution $\phi_m$, from the suite of MMS simulations. The solid lines represent the expected convergence rate for each order.}
      \label{fig:mms_convergence_analysis}
   \end{center}
\end{figure}

Figure \ref{fig:mms_phi} demonstrates how $\phi$ converges towards the manufactured solution $\phi_m$ as the grid is refined. The convergence rate for each order of the central difference scheme is illustrated in Figure \ref{fig:mms_convergence_analysis}. The anomaly in the twelfth-order convergence plot was likely caused by reaching the limit of machine precision. Overall, these results provide confidence in the correctness of the automatically-generated code/model.

\subsection{3D Taylor-Green vortex}\label{sect:taylor_green_vortex}
The Taylor-Green vortex is a well-known hydrodynamic problem \citep{Brachet_etal_1983, DeBonis_2013, BullJameson_2014} characterised by transition to turbulence, decay of turbulence, and the energy dissipation during its evolution. It is frequently used to evaluate the ability of a numerical method to capture the underlying physical processes. During the initial stages of evolution, the dynamics display structural changes (rolling up, streching and interaction of the vortices). This process is inviscid in nature. Later the vortices break down and transition into fully-turbulent dynamics. As there are no external forces or turbulence-generating mechanisms, the small-scale structures dissipate all the energy, and the fluid eventually comes to rest \citep{Brachet_etal_1983}. The numerical method employed should be able to capture each of these stages accurately.

The 3D compressible Navier-Stokes equations were solved in non-dimensional form, written in Einstein notation as

\begin{equation}\label{eq:mass}
   \frac{\partial \rho}{\partial t} + \frac{\partial}{\partial x_j}\left[\rho u_j\right] = 0,
\end{equation}
\begin{equation}\label{eq:momentum}
   \frac{\partial \rho u_i}{\partial t} + \frac{\partial}{\partial x_j}\left[\rho u_i u_j + p\delta_{ij} - \tau_{ij}\right] = 0,
\end{equation}
and

\begin{equation}\label{eq:energy}
   \frac{\partial \rho E}{\partial t} + \frac{\partial}{\partial x_j}\left[\rho E u_j + u_j p - q_j - u_i\tau_{ij}\right] = 0.
\end{equation}
for the conservation of mass, momentum and energy, respectively. The (dimensionless) quantity $\rho$ is the fluid density, $u_i$ is the $i$-th (scalar) component of the velocity vector $\mathbf{u}$, $p$ is the pressure field, $E$ is the total energy. The components of the stress tensor $\tau$ are given by

\begin{equation}
   \tau_{ij} = \frac{1}{\mathrm{Re}}\left(\frac{\partial u_i}{\partial x_j} + \frac{\partial u_j}{\partial x_i} - \frac{2}{3}\delta_{ij}\frac{\partial u_k}{\partial x_k}\right),
\end{equation}
where $\delta_{ij}$ is the Kronecker Delta function and $\mathrm{Re}$ is the Reynolds number. The components of the heat flux term $q$ are given by

\begin{equation}
q_j = \frac{\mu}{(\gamma-1)\ \mathrm{M}^2\ \mathrm{Pr}\ \mathrm{Re}}\frac{\partial T}{\partial x_j},
\end{equation}
where $T$ is the temperature field, $\gamma$ is the ratio of specific heats, $\mathrm{M}$ is the Mach number, and $\mathrm{Pr}$ is the Prandtl number. The various quantities are non-dimensionalised using the reference velocity $u_{\mathrm{ref}}$, the reference length $L$, the reference density $\rho_{\mathrm{ref}}$, and the reference temperature $T_{\mathrm{ref}}$.

The equation of state linking $p$, $\rho$ and $T$, is defined by

\begin{equation}\label{eq:eos}
   p = \frac{1}{\gamma \mathrm{M}^2} \rho T,
\end{equation}
and the total energy is given by

\begin{equation}
   \rho E = \frac{p}{\gamma - 1} + \frac{1}{2}\rho u_j^2.
\end{equation}
The pressure $p$ is non-dimensionalised by $\rho_{\mathrm{ref}}u_{\mathrm{ref}}^2$.

Central finite difference schemes are non-dissipative and are therefore suitable for accurately capturing turbulent dynamics. However, the lack of dissipation can make the scheme unstable. To improve the stability, a skew-symmetric formulation \citep{BlaisdellMansourReynolds_1991, BlaisdellMansourReynolds_1993, BlaisdellSpyropoulosQin_1996, Pirozzoli_2011} was applied to the convective terms in (\ref{eq:mass}), (\ref{eq:momentum}) and (\ref{eq:energy}); the convective term then becomes

\begin{equation}\label{skew_symmetric_formulation}
\frac{\partial}{\partial x_j}[\rho \phi u_{j}] = \frac{1}{2} \left(\frac{\partial}{\partial x_j}\rho \phi u_{j} + u_{j} \frac{\partial}{\partial x_j}\rho \phi  + \rho \phi \frac{\partial}{\partial x_j}u_{j} \right),
\end{equation}
where $\phi$ should be set to $1$, $u_j$ and $E$ for the continuity, momentum and energy equations, respectively. It should also be noted that the both the convective and viscous terms are discretised using the same spatial order. In all of the simulations performed, the Laplacian in the viscous term is expanded using a finite difference representation of the second derivative (i.e.~not treated by successive first derivatives).

As per the work of \cite{DeBonis_2013} and \cite{BullJameson_2014}, the equations were solved in a 3D cube, with $0 \leq x_0 \leq 2\pi L$, $0 \leq x_1 \leq 2\pi L$, and $0 \leq x_2 \leq 2\pi L$. Periodic boundary conditions were applied on all surfaces. The following initial conditions were imposed at time $t$ = 0:

\begin{equation}
   u_0(x_0, x_1, x_2, t=0) = \sin\left(\frac{x_0}{L}\right)\cos\left(\frac{x_1}{L}\right)\cos\left(\frac{x_2}{L}\right),
\end{equation}
\begin{equation}
   u_1(x_0, x_1, x_2, t=0) = -\cos\left(\frac{x_0}{L}\right)\sin\left(\frac{x_1}{L}\right)\cos\left(\frac{x_2}{L}\right),
\end{equation}
\begin{equation}
   u_2(x_0, x_1, x_2, t=0) = 0,
\end{equation}
\begin{equation}
   p(x_0, x_1, x_2, t=0) = \frac{1}{\gamma \mathrm{M}^2} + \frac{1}{16}\left(\cos\left(\frac{2x_0}{L}\right)+\cos\left(\frac{2x_1}{L}\right)\right)\left(2 + \cos\left(\frac{2x_2}{L}\right)\right),
\end{equation}
In all the simulations, $\mathrm{Re}$ = 1,600, $\mathrm{Pr}$ = 0.71, $\mathrm{M}$ = 0.1, and $\gamma$ = 1.4. The reference quantities $L$, $u_{\mathrm{ref}}$ and $\rho_{\mathrm{ref}}$ were set to 1.0, and the reference temperature $T_{\mathrm{ref}}$ was evaluated using the equation of state (\ref{eq:eos}).

A fourth-order accurate central differencing scheme was used to spatially discretise the domain, and a third-order Runge-Kutta timestepping scheme was used to march the equations forward in time. A set of simulations was performed over a range of resolutions, namely $64^3$, $128^3$, $256^3$ and $512^3$ uniformly-spaced grid points. For the 64$^3$ case, a non-dimensional time-step size $\Delta t$ of 3.385 $\times$ 10$^{-3}$ \cite{DeBonis_2013} was used. Each time the number of grid points was doubled, the time-step size was halved to maintain a constant upper bound on the Courant number. The generated code was targetted towards the CUDA backend using OPS and executed on an NVIDIA Tesla K40 GPU until a non-dimensional time of $t$ = 20, except for the 512$^3$ case; this was targetted towards the MPI backend and run in parallel over 1,440 processes on the UK National Supercomputing Service (ARCHER) due to lack of available memory on the GPU, and provided a good example of how the backend can be readily changed.

\begin{figure}[!ht]
   \begin{center}
      \includegraphics[width=0.49\columnwidth]{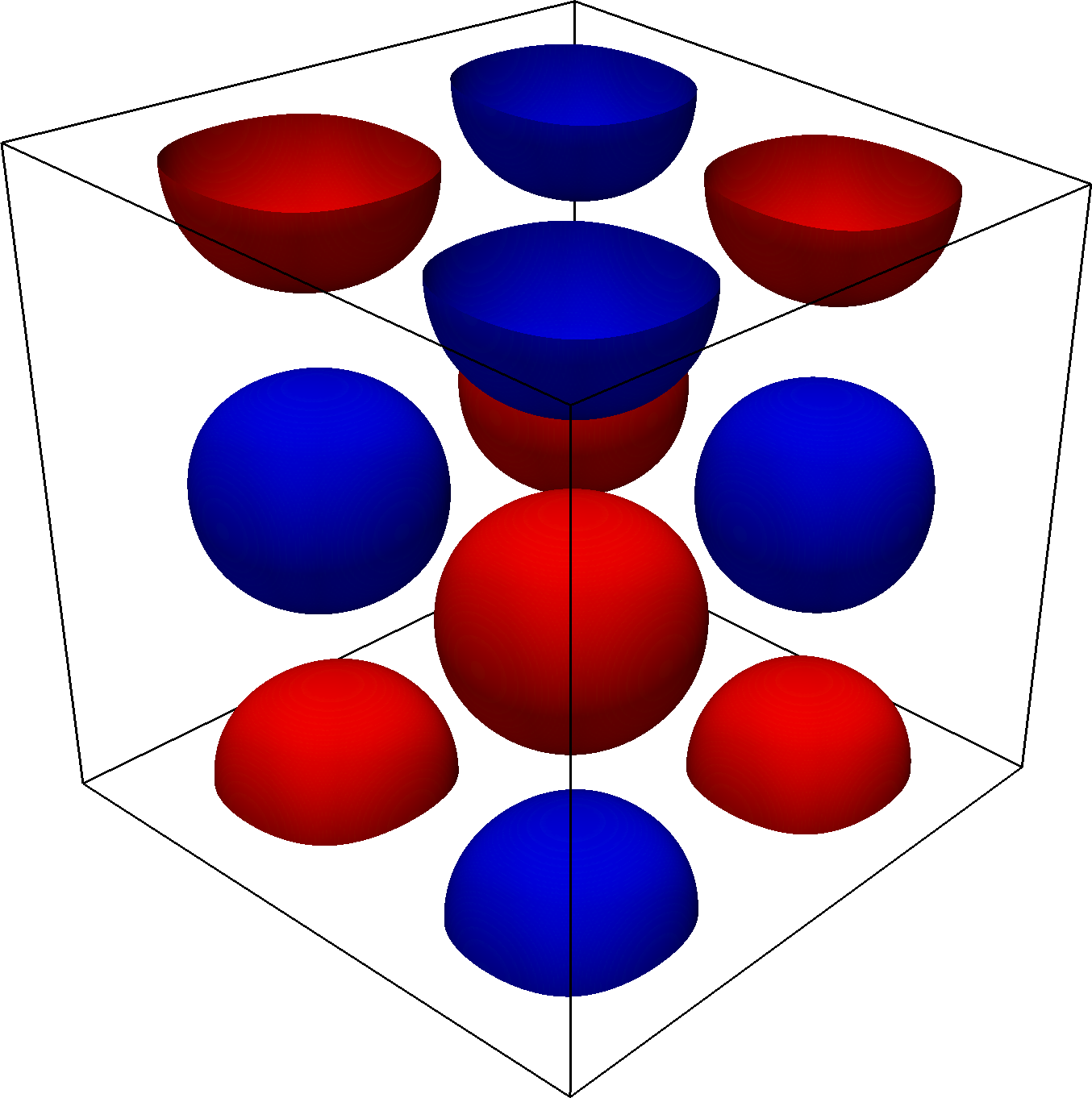}
      \includegraphics[width=0.49\columnwidth]{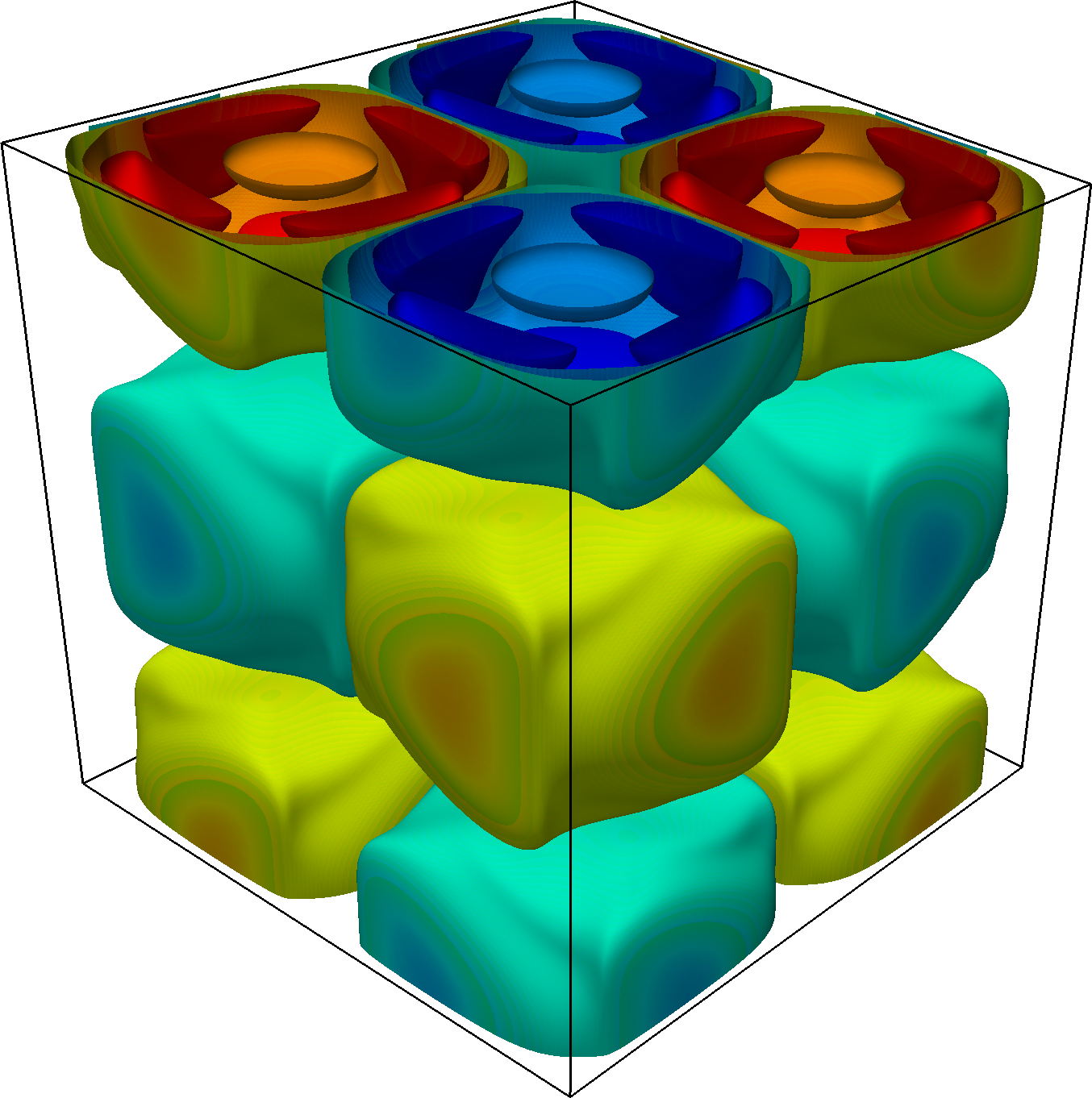}\\
      \includegraphics[width=0.49\columnwidth]{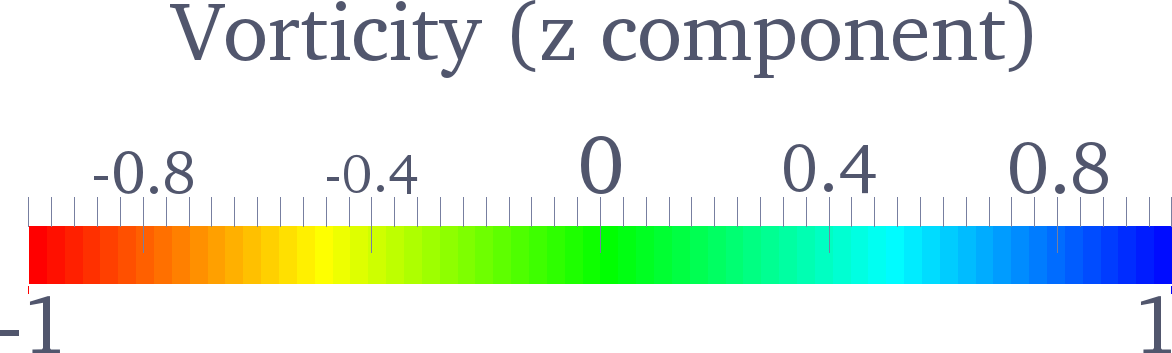}
      \includegraphics[width=0.49\columnwidth]{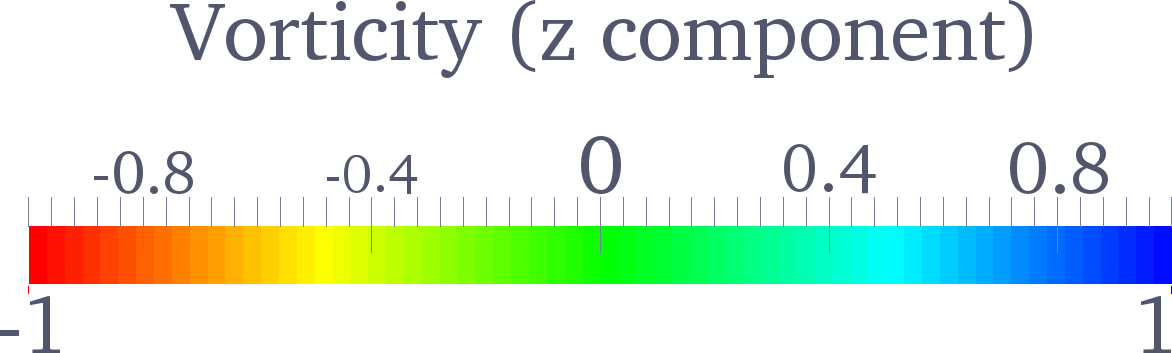}\\
      \includegraphics[width=0.49\columnwidth]{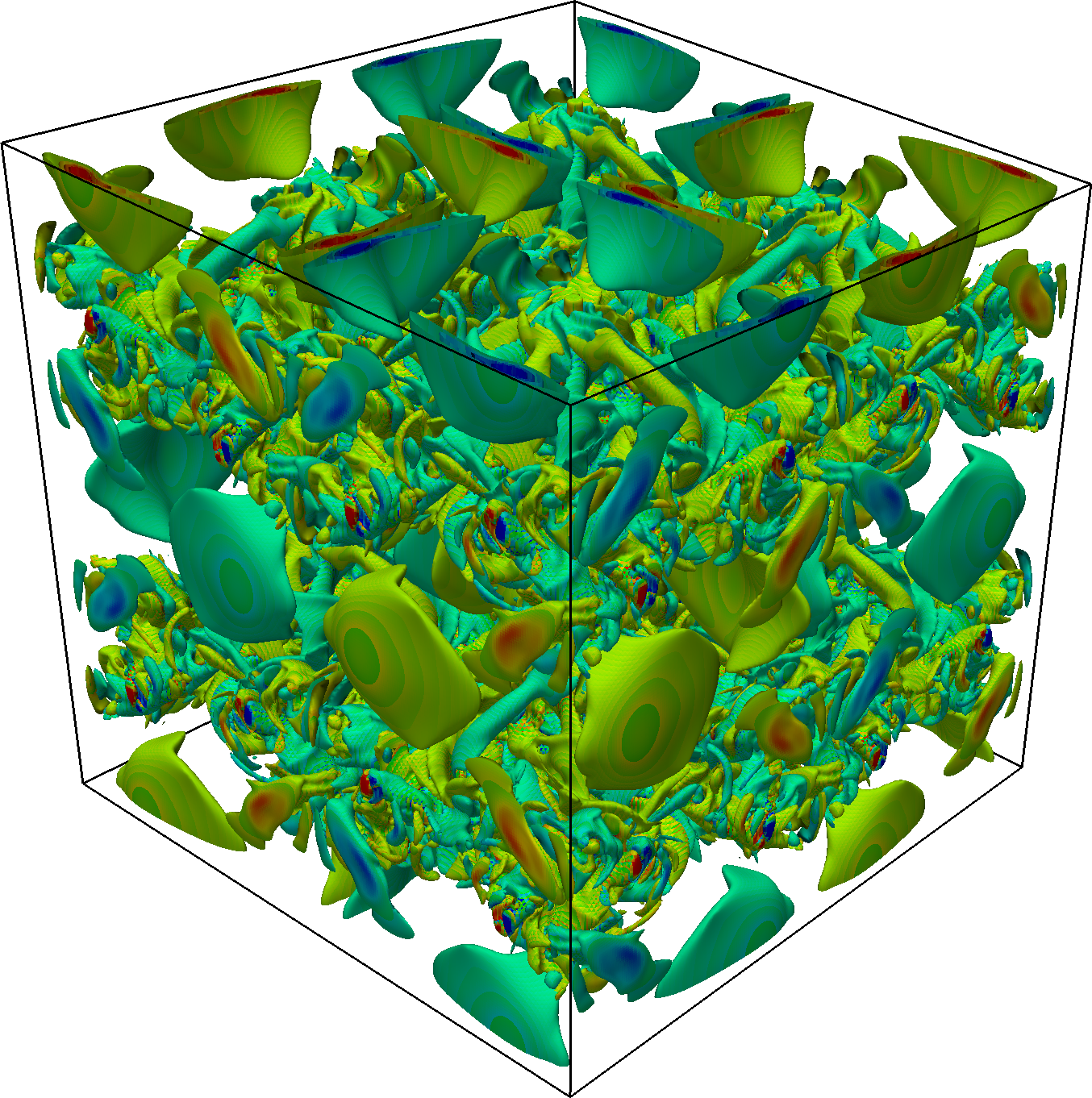}
      \includegraphics[width=0.49\columnwidth]{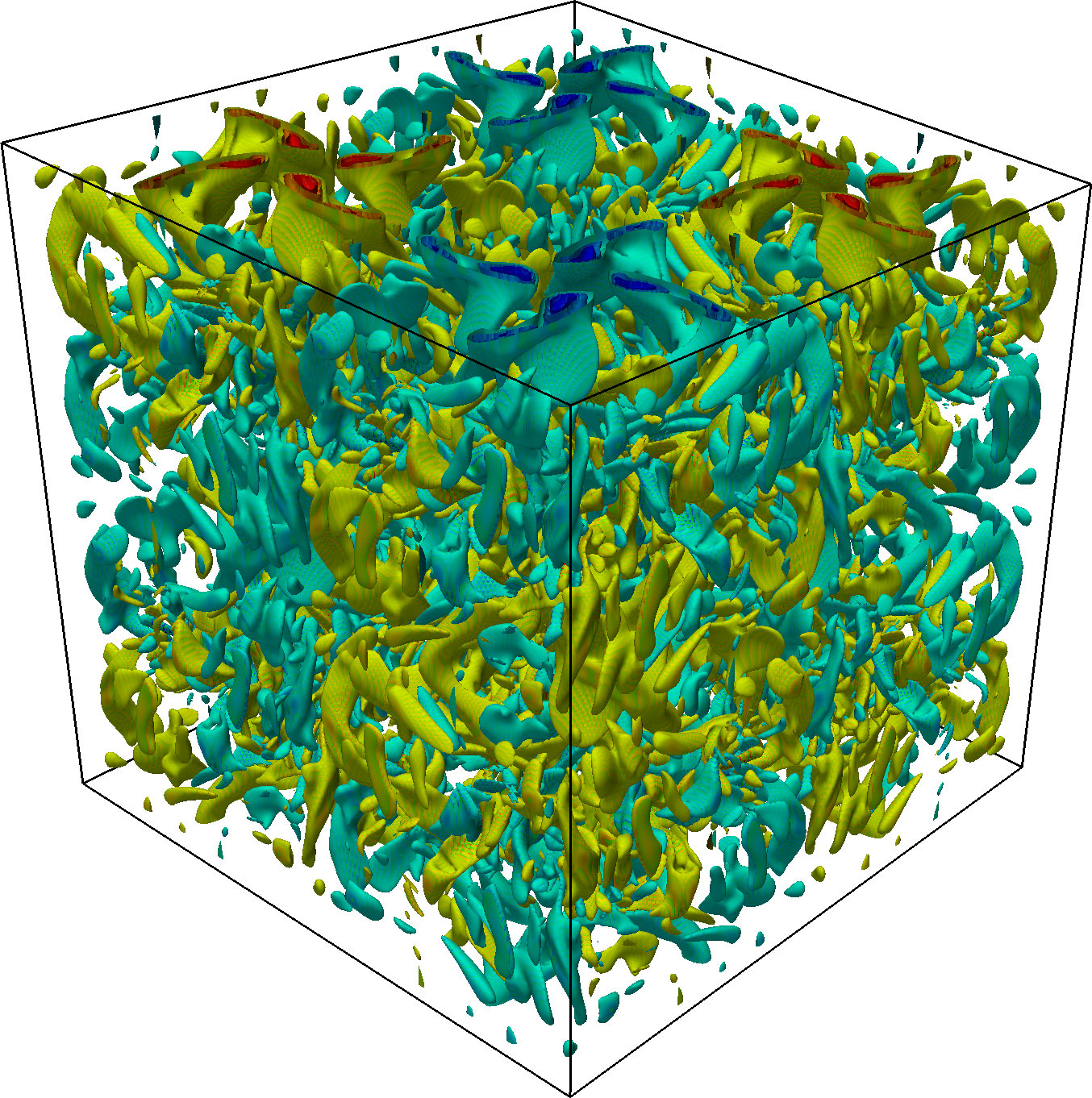}\\
      \includegraphics[width=0.49\columnwidth]{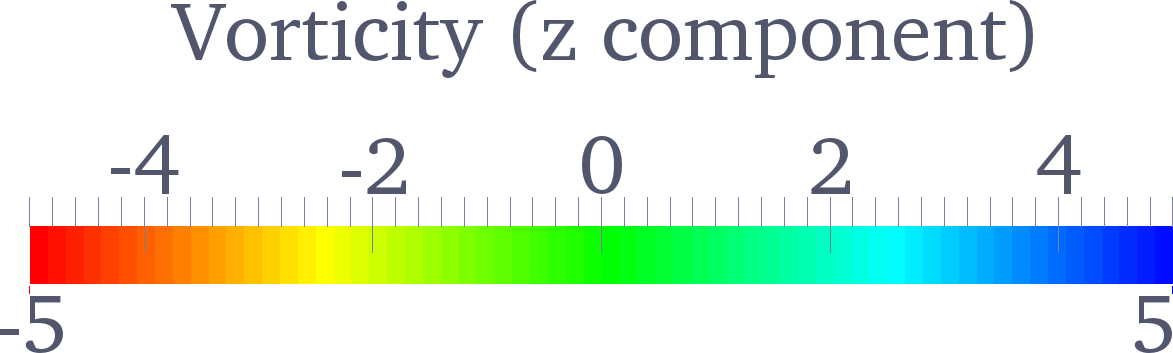}
      \includegraphics[width=0.49\columnwidth]{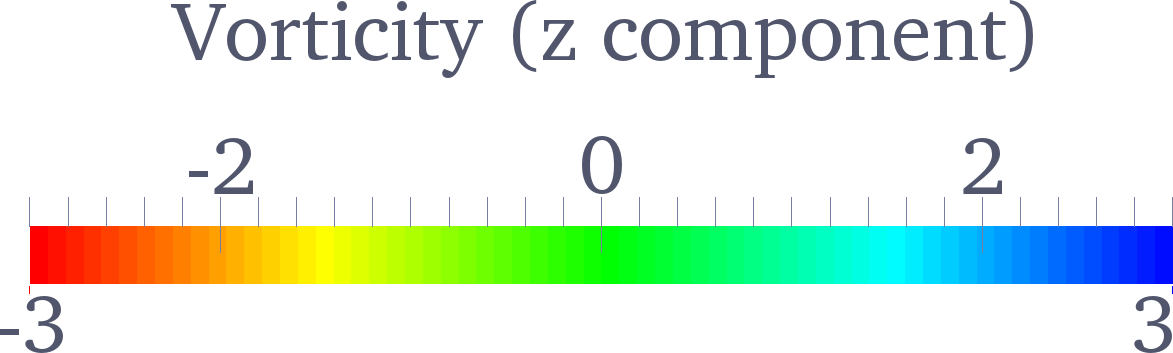}
      \caption{Visualisations of the non-dimensional vorticity ($z$-component) iso-contours, from the Taylor-Green vortex test case with a 256$^3$ grid, at various non-dimensional times. Top left to bottom right: non-dimensional time $t$ = 0, 2.5, 10, 20.}
      \label{fig:tgv_z_vorticity}
   \end{center}
\end{figure}

The $z$-component of the vorticity field at various times can be found in Figure \ref{fig:tgv_z_vorticity}. At non-dimensional time $t$ = 2.5 vortex evolution and stretching are clearly visible, progressing onto highly turbulent dynamics where the relatively smooth structures roll-up and eventually breakdown at around $t$ = 9. This point is characterised by peak enstrophy in the system. The final stage of the simulation features the decay of the turbulent structures such that the enstrophy tends towards its initial value.

\begin{figure}[!ht]
   \begin{center}
      \includegraphics[width=\columnwidth]{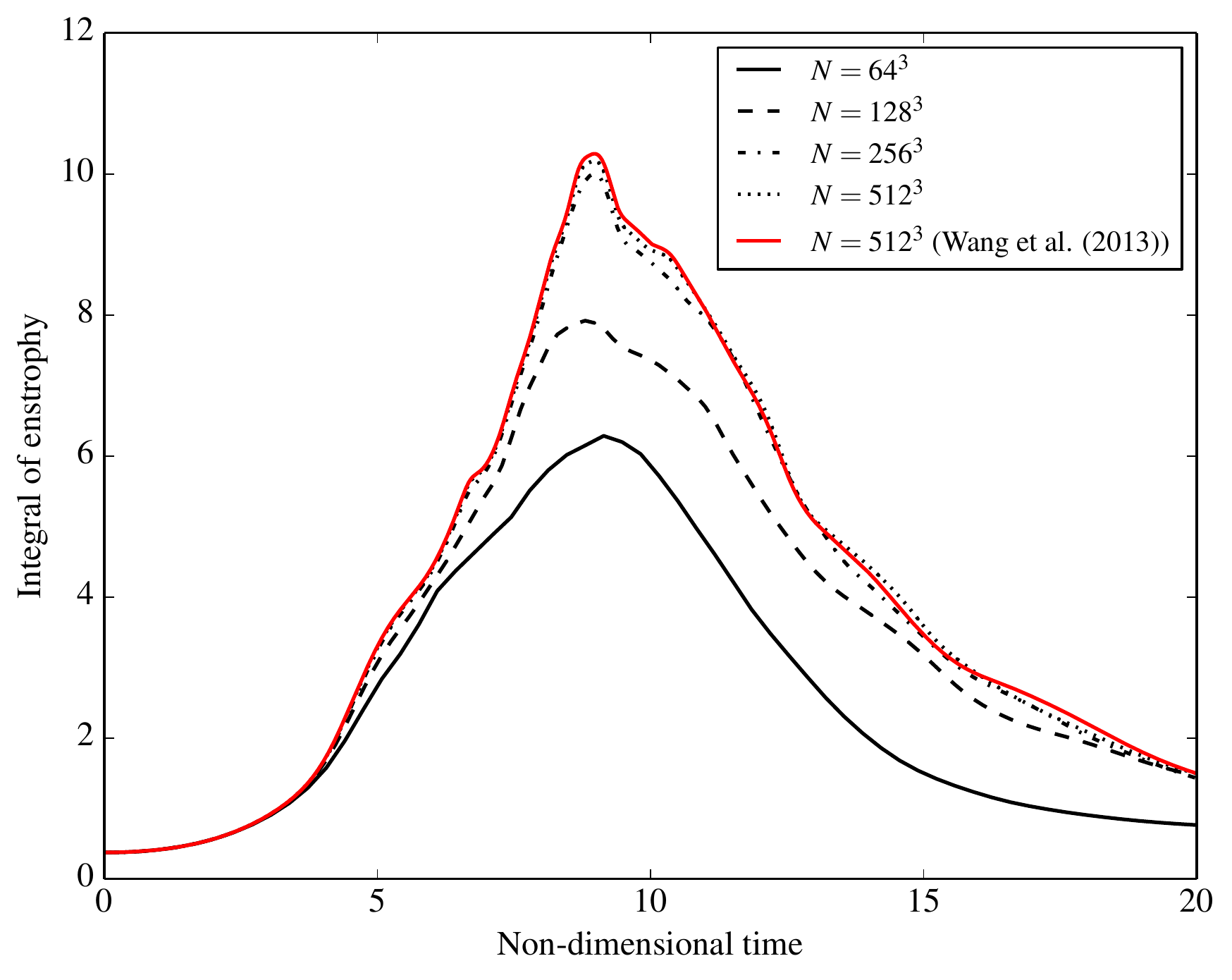}
      \caption{The integral of the enstrophy in the domain until non-dimensional time $t$ = 20, from the Taylor-Green vortex test case. The reference data from \cite{Wang_etal_2013} is also shown for comparison.}
      \label{fig:tgv_enstrophy}
   \end{center}
\end{figure}

\begin{figure}[!ht]
   \begin{center}
      \includegraphics[width=\columnwidth]{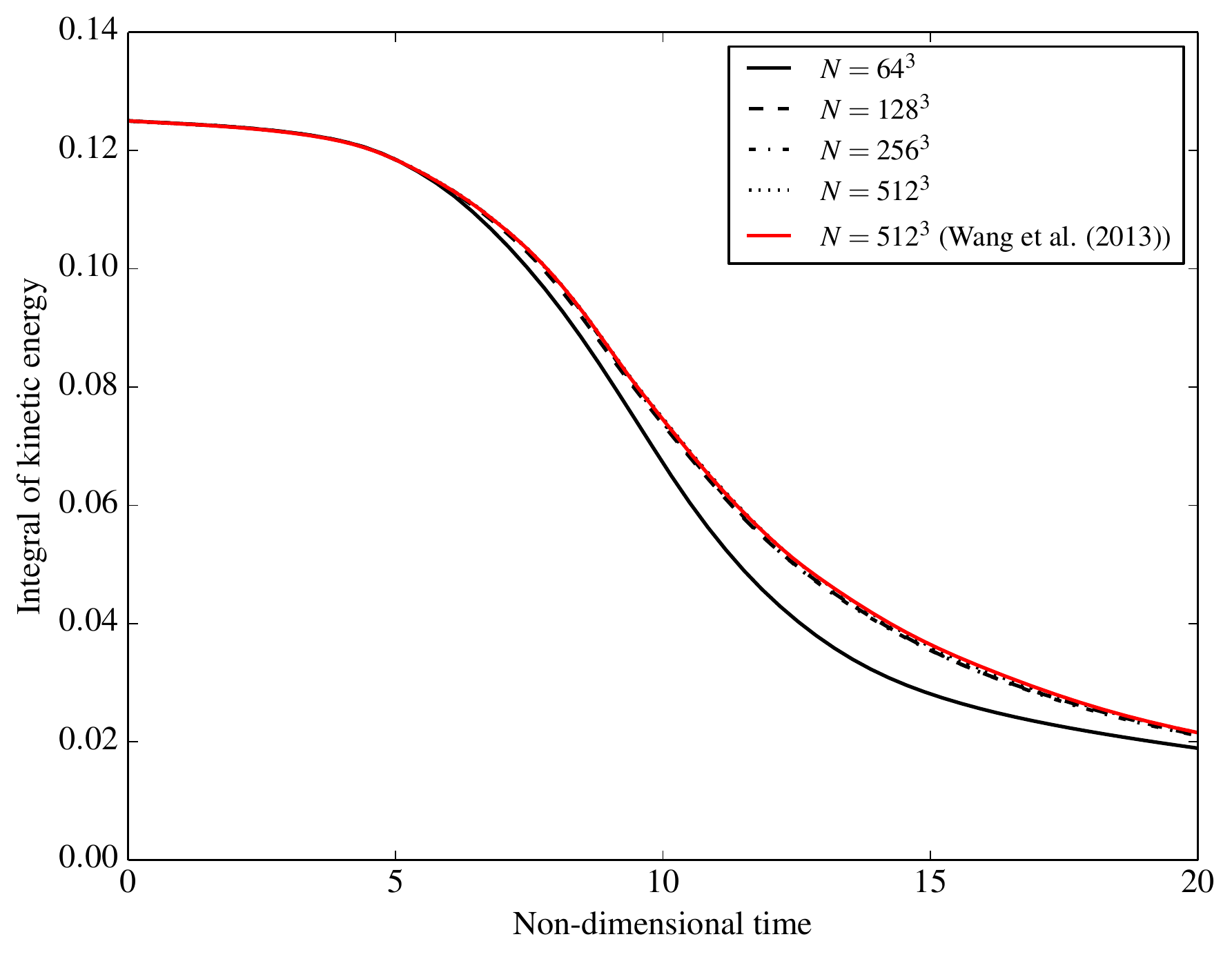}
      \caption{The integral of the kinetic energy in the domain until non-dimensional time $t$ = 20, from the Taylor-Green vortex test case. The reference data from \cite{Wang_etal_2013} is also shown for comparison.}
      \label{fig:tgv_ke}
   \end{center}
\end{figure}

Following the definitions of \cite{DeBonis_2013}, the integrals of the kinetic energy

\begin{equation}
   E_k = \frac{1}{\rho_{\mathrm{ref}}\Omega}\int_{\Omega}\frac{1}{2}\rho u_j u_j\ \mathrm{d}\Omega,
\end{equation}
and enstrophy

\begin{equation}
   \varepsilon = \frac{1}{\rho_{\mathrm{ref}}\Omega}\int_{\Omega}\frac{1}{2}\rho \left(\epsilon_{ijk}\frac{\partial u_k}{\partial x_j}\right)^2\ \mathrm{d}\Omega,
\end{equation}
were computed throughout the simulations. Note that $\Omega$ is the whole domain and $\epsilon_{ijk}$ is the Levi-Civita function. These quantities are shown in Figures \ref{fig:tgv_enstrophy} and \ref{fig:tgv_ke} for the various grid resolutions, and are plotted against the reference data from a spectral element simulation by \cite{Wang_etal_2013} using a 512$^3$ grid for comparison. Figure \ref{fig:tgv_ke} highlights the inviscid nature of the Taylor-Green vortex problem for $t$ $<$ $\sim$3-4. The transition to turbulence occurs from $\sim$3$<t<$9 (which is associated with the peak in enstrophy in Figure \ref{fig:tgv_enstrophy}). Finally, dissipation occurs at $t$ $>$ 9. The results show a clear agreement with the reference data, and represents a solid first step towards the validation of OpenSBLI.

\section{Conclusion}\label{sect:conclusion}
Advances in compute hardware are driving a need to change the current state of numerical model development. By developing a new modelling framework based on automated solution techniques, we have effectively future-proofed the core of the SBLI codebase; no longer does a computational scientist need to re-write significant portions of code in order to get it up and running on a new piece of hardware. Instead, the model is derived from a high-level specification independent of the architecture that it will run on, and the underlying code is automatically generated and tailored to a particular backend, the responsibility for which would rest with computer scientists who are experts in parallel programming paradigms. Furthermore, the ease at which the governing equations can be changed is a fundamental advantage of using such abstract specifications. This was highlighted here by considering three test cases, each of which comprised a different set of equations. The discretisation, code generation and code targetting is performed automatically, thereby reducing development costs and potentially avoiding errors, bugs, and non-performant/non-optimal operations. In addition, code that solves the different variants of the same governing equations can be easily generated. For example, in the compressible Navier-Stokes equations, viscosity can be treated either as a constant or as a spatially-varying term. In static, hand-written codes this flexibility comes at the cost of writing different routines for the various formulations, unlike with automated code generation techniques. This is particularly useful when wanting to switch between Cartesian and generalised coordinates. This particular framework also facilitates the fast and efficient switching between different spatial orders of accuracy, and reduces the development time and effort when wishing to try out new numerical formulations of the equations (or a new spatial/temporal scheme) on a wide variety of test cases.

\subsection{Future work}
Explicit schemes such as the one implemented here can be readily extendible to a range of application areas such as computational aeroacoustics, aero-thermodynamics, problems involving shocks, and hypersonic flow. Incompressible flows may also be handled with the explicit, compressible solver in OpenSBLI so long as the Mach number is sufficiently small. However, this puts tight restrictions on the time-step size thereby limiting the efficiency of the solver, and thus limits the range of applications that OpenSBLI can handle within the context of CFD in general. 

Extending to implicit timestepping schemes requires backend support from OPS for matrix inversion, for example. Once this support is implemented, extending OpenSBLI to handle implicit timestepping schemes is straightforward.

The treatment of incompressible flows can be accomplished using schemes such as pressure projection methods \cite{Chorin_1967, Chorin_1968}. Such a method requires (1) the solution of a tentative velocity field, (2) the solution to a pressure Poisson equation (using either direct or iterative solvers), and (3) the update/correction of the velocity field. The equations defining each step would need to be given by the user in OpenSBLI, in a similar fashion to implementing a projection method in the Unified Form Language (UFL) in FEniCS \cite{FEniCS_2011, Alnaes_etal_2014}. OpenSBLI would also need to recognise that a projection method has been chosen, possibly via a flag set in the problem definition file. For step (2) of the method, direct solvers can be implemented directly once support for matrix inversion and fast Fourier transforms (for example) are included in OPS (which in turn would need to link to various linear algebra packages such as PETSc \cite{Balay_etal_2014}). This is similar to how an implicit time-stepping scheme would be implemented in OpenSBLI. On the other hand, explicit solution schemes require an iterative solution to the pressure Poisson equation; this is possible by writing the relevant kernel support with an exit criterion (which exits the kernel once a desired tolerance for the solution residual has been attained) and modifying how the code is generated with this in mind. Other methods for incompressible flows such as artificial compressibility methods \cite{Chorin_1967, Chorin_1967_2, OhwadaAsinari_2010} can be implemented with the current functionality by modifying the input equations in the problem setup file accordingly.

The work considered here only focussed on the MPI and CUDA backends for CPU and GPU execution, respectively. Future work will consider the CPU and GPU performance on other backends, such as OpenMP. For problems such as Mandelbrot Set computation and matrix multiplication, OpenMP has been demonstrated to perform well against other APIs such as CUDA and OpenACC \cite{Ledur_etal_2013}. Future work will also look at comparing the performance of the legacy Fortran-based SBLI code against the OpenSBLI-generated code in order to demonstrate the potential speed-ups that can be obtained.

\section{Code Availability}\label{sect:code_availability}
OpenSBLI is an open-source release of the original SBLI code developed at the University of Southampton, and is available under the GNU General Public Licence (version 3). Prospective users can download the source code from the project's Git repository: \texttt{https://github.com/opensbli/opensbli}

\section{Acknowledgments}
CTJ was supported by a European Union Horizon 2020 project grant entitled ``ExaFLOW: Enabling Exascale Fluid Dynamics Simulations'' (grant reference 671571). SPJ was supported by an EPSRC grant entitled ``Future-proof massively-parallel execution of multi-block applications'' (EP/K038567/1). The data behind the results presented in this paper will be available from the University of Southampton's institutional repository. The authors acknowledge the use of the UK National Supercomputing Service (ARCHER), with computing time provided by the UK Turbulence Consortium (EPSRC grant EP/L000261/1). The authors would also like to thank the NVIDIA Corporation for donating the Tesla K40 GPU used throughout this research.

\appendix

\section{Example of a simulation setup file}\label{sect:setup}

\begin{figure}[!ht]
   \begin{center}
      \includegraphics[width=\columnwidth]{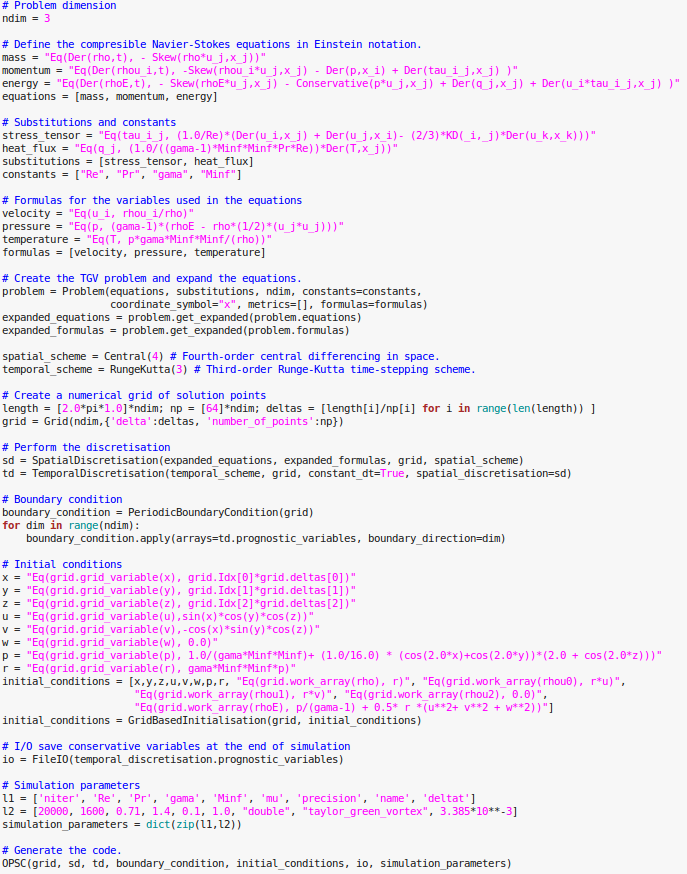}
      \caption{A cut-down version of the 3D Taylor-Green vortex setup/configuration file (67 lines long including whitespace), showing the key components and classes available in OpenSBLI.}
      \label{fig:setup}
   \end{center}
\end{figure}

The code in Figure \ref{fig:setup} contains the key components of a simulation setup file. Specifically, this is taken from the Taylor-Green vortex simulation. Other examples of setup files can be found in the \texttt{apps} directory of OpenSBLI.

\section*{References}
\bibliographystyle{elsarticle-num}
\bibliography{opensbli}







\end{document}